# Impact of Third Order Dispersion on Dissipative Soliton Resonance

Lujun Hong, Yazhou Wang, Abubakar Isa Adamu, Md. Selim Habib, Mengqiang Cai, Jie Xu, Jianfeng Li, and Christos Markos

*Abstract*—Dissipative soliton resonance (DSR) is a promising way for high-energy pulse generation typically having a symmetrical square pulse profile. While this method is well known, the impact of third order dispersion (TOD) on DSR is yet to be fully addressed in the literature. In this article, the impact of TOD on DSR is numerically investigated under the frame of the complex cubic-quintic Ginzburg-Landau equation (CQGLE). Our numerical investigations indicate that DSR can stably exist under TOD with nearly the same pulse amplitude, but with a (significantly) different pulse duration. Depending on the value of chromatic dispersion, the pulse duration can be notably longer or shorter due to the presence of TOD. The TOD effect also alters the dependence of pulse duration on the nonlinear gain. Another impact of TOD on DSR is that the DSR exists with an asymmetric pulse profile, leading to steepening of one edge of the DSR pulse, while flattening of the other. Our results indicate that TOD has a critical role for realizing DSR in mode-locked lasers and it should be taken into consideration during design and development of DSR-based lasers.

## I. INTRODUCTION

DISSIPATIVE soliton resonance (DSR) effect has attracted wide attention since it was first predicted as a promising way for high energy pulse generation in 2008 [1]–[6]. Quite different from other kinds of stationary pulse regimes, the DSR shows distinctive advantage of wave-breaking-free effect with a square-shaped pulse characterized by flat-top and steep pulse edges, where the pulse energy infinitely increases through expanding pulse duration without altering the pulse amplitude [7]–[11]. Not only the unlimited pulse energy is attractive, but also its distinct square pulse shape is promising for applications such as the efficient optical parametric oscillators with square pulse pump [12]–[15]. Near linear chirp distribution across the flat-top region of DSR allows for the efficient compression of pulse duration outside the laser oscillator. Owing to these advantages, the formation mechanism and existence condition of DSR have been thoroughly discussed numerically and theoretically [7], [16], [17], inspiring a number of experimental reports in the frame of mode-locked fiber lasers [2], [5], [6]. CQGLE) as a phenomenological description of mode-locking dynamics averages the effects of discrete laser cavity components over one round trip [18], and thus being widely adopted to explore the universal behavior of DSR. With this model, it turns out that DSR can be stably formed in both anomalous and normal chromatic dispersion regimes[7], [19], [20], and remains robust in laser models with parameter management [21]. The numerical investigation of DSR in this framework also guides the practical design of DSR by relating the parameters of the CQGLE with realistic ring laser cavity [22].

In the aforementioned investigations, the DSR has been studied however without considering the effect from high order dispersion (HOD). Particularly, mode-locked fiber laser with a long cavity length could easily accumulate a high amount of TOD , which can result in asymmetry of pulse profile [23], [24], pulse broadening or compression [25], [26], and impair soliton stability [27]. Complex pulse regimes such as soliton pulsating and erupting [28], [29], and even chaotic pulse regimes can be introduced by TOD effect [30]. Appropriate TOD can eliminate soliton explosions through compensating self-steepening effect [31]. Fourth order dispersion has also been investigated in terms of soliton stability, soliton splitting [32], [33], pure-quartic soliton formation [34], etc. Inspired by these complex roles of HOD, a fascinating question arises on how is DSR affected by

This work was supported by National Natural Science Foundation of China (Grant No. 11904152), Start-Up Funding of Nanchang University (9167-28770244), Danmarks Frie Forskningsfond Hi-SPEC project (Grant No. 8022-00091B), Villum Fonden (36063), and Multi-BRAIN project (R276-2018-869) funded by the Lundbeck Foundation. *(Corresponding author: Yazhou Wang; Jianfeng Li)*

Lujun Hong is with the Institute of Space Science and Technology, Nanchang University, Nanchang 330031, China, and with the Department of Photonics Engineering, Technical University of Denmark, DK-2800 Lyngby, Denmark (e-mail: lunhong@dtu.dk).

Yazhou Wang is with the Department of Photonics Engineering, Technical University of Denmark, DK-2800 Lyngby, Denmark, and with the School of Optoelectronic Science and Engineering, State Key Laboratory of Electronic Thin Films and Integrated Devices, University of Electronic Science and Technology of China, Chengdu 610054, China (e-mail: yazwang@fotonik.dtu.dk).

Abubakar Isa Adamu is with Lumenisity Ltd, U.K. (e-mail: abisa@fotonik.dtu.dk).

Md. Selim Habib is with the Department of Electrical and Computer Engineering, Florida Polytechnic University, Lakeland FL-33805 USA (e-mail: mhabib@floridapoly.edu).

Mengqiang Cai is with the Institute of Space Science and Technology, Nanchang University, Nanchang 330031, China (e-mail: caimengqiang@ncu.edu.cn).

Jie Xu is with School of Medical Information and Engineering, Southwest Medical University, Luzhou 646000, China (e-mail: xujie011451@163.com).

Jianfeng Li is with the School of Optoelectronic Science and Engineering, State Key Laboratory of Electronic Thin Films and Integrated Devices, University of Electronic Science and Technology of China,Chengdu 610054,China (e-mail: lijianfeng@uestc.edu.cn).

Christos Markos is with the Department of Photonics Engineering, Technical University of Denmark, DK- 2800 Lyngby, Denmark, and also with the NORBLIS ApS, 2830 Virum, Denmark (e-mail: chmar@fotonik.dtu.dk).

HOD effect. If the pulse profile of DSR can be reshaped without impairing its wave-breaking feature, it will provide higher freedom for potential applications.

In this paper, we thoroughly investigate the performance of DSR in the frame of CQGLE by introducing TOD effect. The simulation result shows that, depending on the amount and sign of chromatic dispersion, the TOD effect can lead to a significant expansion or reduction of the pulse duration of DSR. Besides, TOD also alters the pulse edge steepness of DSR.

## II. NUMERAICAL MODEL

The CQGLE with TOD can be expressed as in [7], [29], [31], [35]

$$i\psi_z + \frac{D}{2}\psi_{tt} - i\beta_3\psi_{ttt} + |\psi|^2\psi + \nu|\psi|^4\psi \\ = i\delta\psi + i\beta\psi_{tt} + i\varepsilon|\psi|^2\psi + i\mu|\psi|^4\psi \quad (1)$$

where the normalized optical envelope $\psi = \psi(t,z)$ is a complex function of retarded time $t$ and propagation distance $z$. $z$ is normalized to the cavity round-trip number. $\psi_z$ is the first-order $z$-derivative, and $\psi_{tt}$ and $\psi_{ttt}$ are the second-order and three-order $t$-derivatives, respectively. Left side of (1) contains the conservative terms: $D$ is the averaged cavity chromatic dispersion, with $D > 0$ ($D < 0$) in the anomalous (normal) chromatic dispersion regime, respectively. $\beta_3$ is TOD of laser cavity, and the sign of $\beta_3$ is consistent with sign of cavity TOD. $\nu$ accounts for the quintic reactive Kerr nonlinearity. Right side of (1) represents all the dissipative effects: $\delta < 0$ is the linear loss. $\beta > 0$ donates spectral filtering. $\varepsilon$ represents the nonlinear gain. $\mu < 0$ counts for saturation of the nonlinear gain.

The complex equation in (1) was numerically solved using the split-step Fourier method. Note that the performance of the DSR is strongly related to the seven parameters. As an example, the parameters used in the simulation are implemented based on the parameters in Ref. [7]: where DSR exists around the upper boundary of the stable soliton solution in the two dimensional parameter space ($D$ and $\varepsilon$) by setting four optimal parameters as $\delta = -0.05, \beta = 0.4, \nu = -0.08,$ and $\mu = -0.05$. In this work, the TOD effect is considered and we focus on analyzing the performance of DSR in the remaining three dimensional parameter space ($\beta_3$, $D$, $\varepsilon$).

## III. SIMULATION RESULTS

### A. Impact of $\beta_3$ on DSR with constant ($D$, $\varepsilon$)

We first investigate the performance of DSR at the point of ($D$=-2.5, $\varepsilon$=0.239677) with $\beta_3 = 0.2$. Stable soliton solution is obtained in this case. Figure 1(a) shows the pulse evolution as a function of round-trip number $z$, where the pulse stably operates with an inclined evolution trace. The slope is proportional to the strength of $\beta_3$. Due to this phenomenon, we move the pulse peak to $t = 0$ when the pulse is close to boundary of time window during the evolution, this ensures the correct simulation within the numerical time window. Figure 1(b) shows the corresponding pulse profile, where the pulse duration and amplitude are $T_{pulse\_duration} = 42.4$ and $|\psi|_{max} = 2.1$, respectively. The pulse profile exhibits a square shape with an approximate flat-top structure (strictly speaking, the flat-top is still a weak convex shape). In order to facilitate comparison, the pulse is translated along lateral axis so that the pulse peak is located at $t = 0$ (the figures below are the same). A typical feature is that both edges of the pulse show different steepness. The rise time, which is defined as the time duration from 5% to 95% of the pulse peak, is used to describe the steepness of pulse edges. In the case of $\beta_3 = 0.2$ (solid line), the rise times of leading and trailing edges of the pulse are $T_{risetime\_leading\_edge} = 12.7$ and $T_{risetime\_traing\_edge} = 23.8$, respectively. As a comparison, Fig. 1(b) also shows the DSR with $\beta_3 = 0$ (the dashed line), where the pulse amplitude is nearly the same for the two pulses. Due to the symmetrical pulse profile at $\beta_3 = 0$, the rise times are $T_{risetime\_leading\_edge} = T_{risetime\_trailing\_edge} = 17.7$, which is between the leading and trailing edges of the DSR with $\beta_3 = 0.2$. This phenomenon suggests that TOD steepens the leading pulse edge but stretches the trailing pulse edge of DSR. In addition, the DSR with zero TOD shows different pulse duration of $T_{pulse\_duration} = 30.0$, indicating the dependence of pulse duration on TOD.

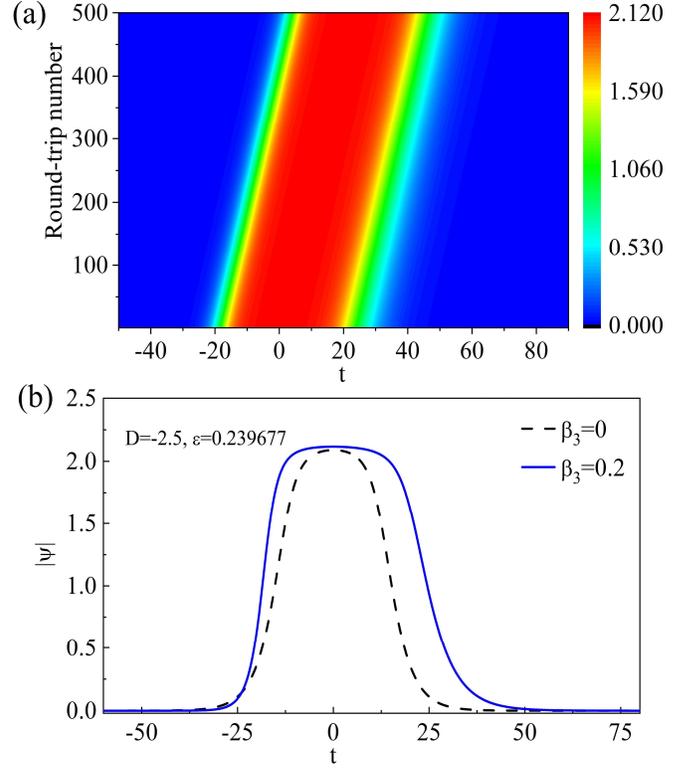

Fig. 1. (a) Evolution of DSR as a function of round-trip number at ($D$=-2.5, $\varepsilon$=0.239677, $\beta_3$=0.2). (b) Pulse profiles for $\beta_3$=0 and $\beta_3$=0.2.

The sign of $\beta_3$ determines if the edge of the pulse is steepened or stretched. The simulation shows that the pulse profile at -$\beta_3$ is always a mirror inversion of the one at $\beta_3$ regarding the axis $t = 0$. For instance, Fig. 2(a) shows the pulse profiles at $\beta_3 = \pm 0.2$, where the minus TOD ($\beta_3 = -0.2$) steepens the trailing edge and stretches the leading edge (the dot-dashed line), and their pulse profiles are mirror symmetric to each other regarding axis $t = 0$. Due to this phenomenon, the following work only considers the cases of $\beta_3 > 0$. Figure 2(b) shows the rise times of leading and trailing pulse edges as a function of $\beta_3$ (the blue solid and dashed lines). The result shows that the TOD steepens the leading pulse edge but flattens the trailing pulse edge. For instance, at the case of $\beta_3 = 0.24$, $T_{risetime\_leading\_edge} = 11.9$, and $T_{risetime\_trailing\_edge} = 25.4$. Figure 2(b) also shows the corresponding variation of pulse duration (the red solid line), and it is found that $T_{pulse\_duration}$ continuously increases from 30.0 to 62.4 as $\beta_3$ goes from 0 to 0.24. Note that, when $\beta_3$ further increases, the pulse duration as it approaches the width of the numerical window and thus cannot be correctly simulated.

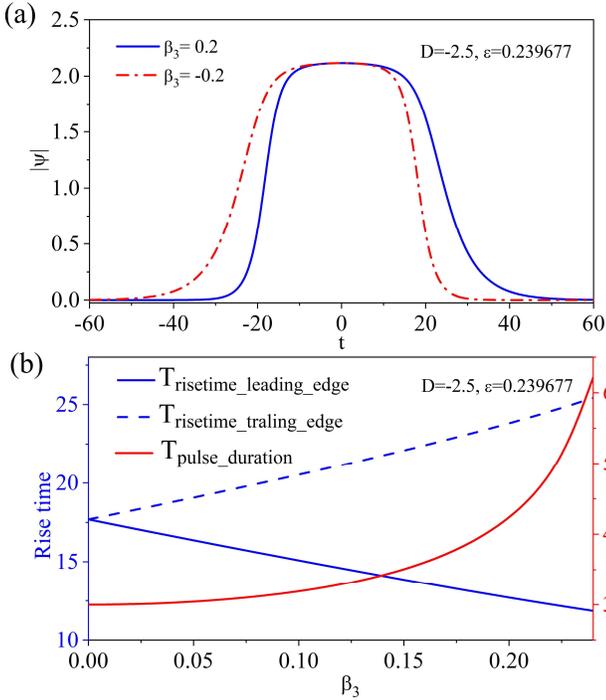

Fig. 2. DSR at ($D$=-2.5, $\varepsilon$=0.239677). (a) Pulse profiles for $\beta_3$=±0.2. (b) Rise times of pulse edges and pulse duration as a function of $\beta_3$.

### B. Rise time of pulse edge in the space of ($\beta_3$, $D$, $\varepsilon$) with constant pulse duration

In fact, the pulse edge steepness depends on not only the TOD ($\beta_3$) but also the chromatic dispersion ($D$) and nonlinear gain ($\varepsilon$). In order to illustrate this, we first simulate the steepness of pulse edges as a function of $D$ without considering TOD. Figure 3(a) shows the rise times of both pulse edges in the range of -2.5<$D$<2.0 with $\beta_3 \neq 0$, where their pulse durations are respectively maintained $T_{pulse\_duration} = 30$, 25, and 20 by appropriately setting the nonlinear gain ($\varepsilon$). It can be seen that, regardless of the pulse duration, the pulse with long pulse edges is formed in large normal dispersion region ($D \ll 0$). As $D$ increases, the rise time continuously decreases to a minimum value of 1.98 at $D \approx 0.3$, and then slowly increases again, indicating that DSR with steepest pulse edges can be obtained in near zero anomalous chromatic dispersion region. In the case of $\beta_3 \neq 0$, the rise time of pulse edges becomes different. For instance, Fig. 3(b) shows the rise times of pulse edges by setting $\beta_3$ to 0.015 for the curve with $T_{pulse\_duration} = 30$ in Fig. 3(a), which is replotted in Fig. 3(b) to facilitate comparison. It can be seen that the leading (trailing) edge is obviously steepened (stretched) in large normal dispersion region. As $D$ increases, the difference of rise times in the cases of with/without TOD becomes weak gradually, and the leading (trailing) pulse edge is stretched (steepened) instead when $D$ is slightly greater than 0 (see inset of Fig. 3(b)).

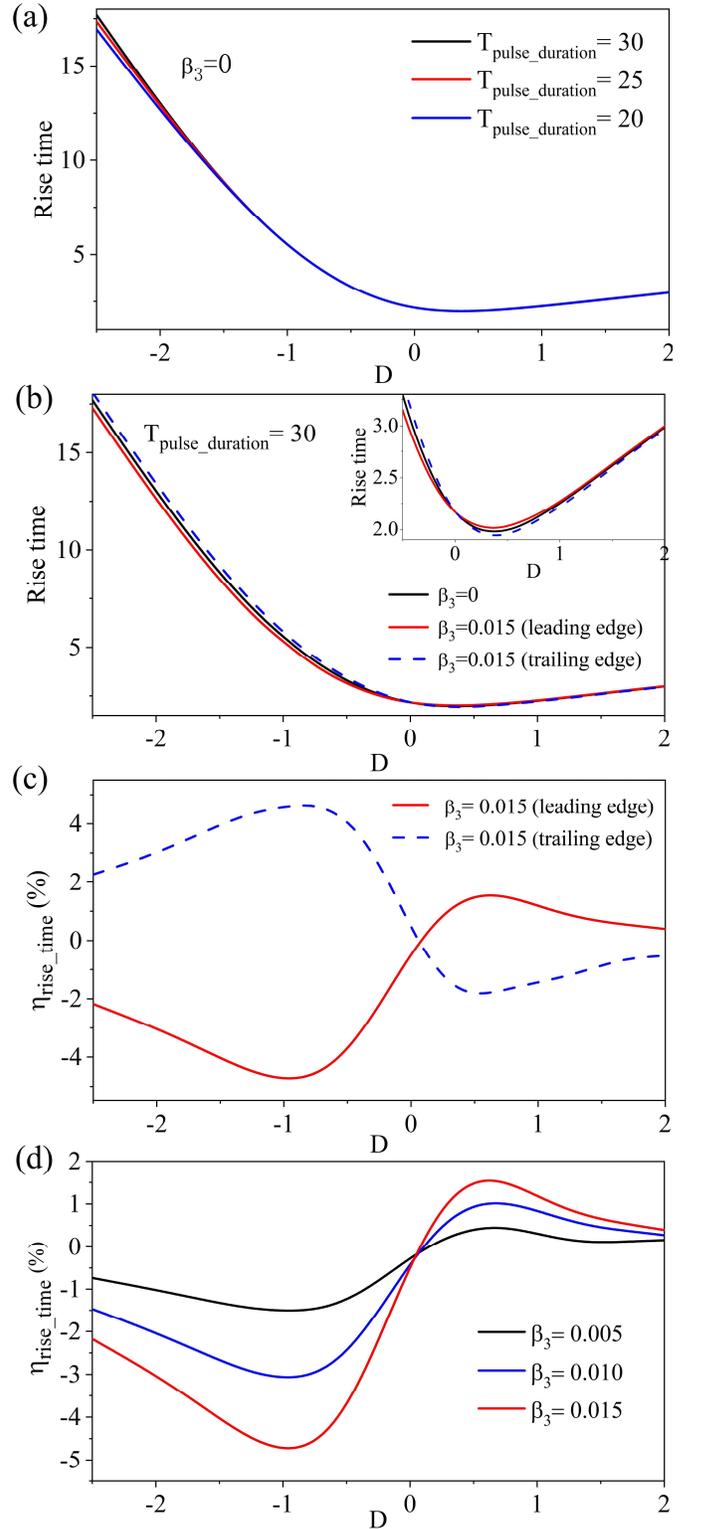

Fig. 3. (a) Rise time of pulse edges as a function of chromatic dispersion with constant pulse duration of $T_{pulse\_duration} = 30$, 25, and 20, respectively. Because $\beta_3$ in (a) is zero, the leading and trailing pulse edges has the same rise time. (b) Rise time of pulse edges with $\beta_3 = 0$ and 0.015, respectively. In (b), the curve with $\beta_3 = 0$ has the fixed pulse duration of $T_{pulse\_duration} = 30$, and the other two curves are calculated based on this curve but with $\beta_3 = 0.015$. (c) $\eta_{rise\_time}$ (the relative change of rise time) corresponds to Fig. 3(b). (d) $\eta_{rise\_time}$ of the leading pulse edge with three different TOD.

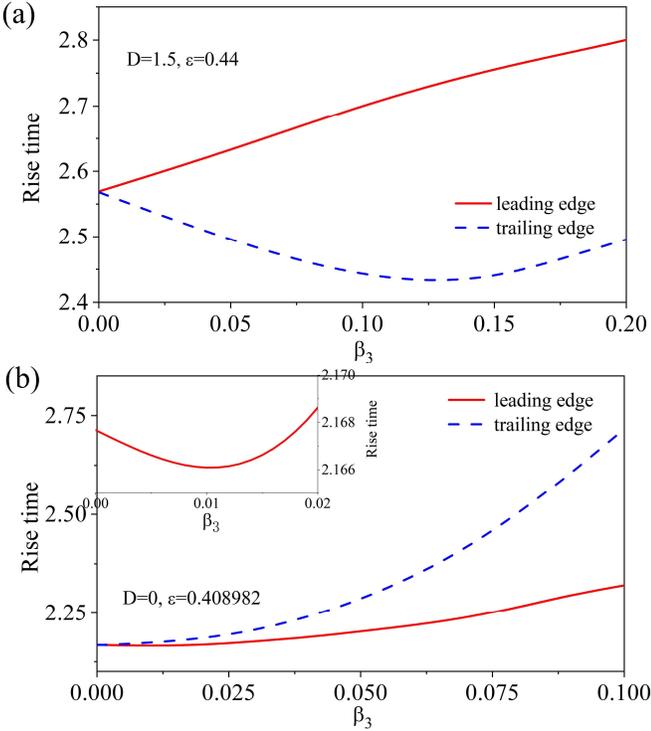

Fig. 4. The rise time of leading and trailing pulse edges as a function of $\beta_3$ at (a) ($D$=1.5, $\varepsilon$=0.44) and (b) ($D$=0, $\varepsilon$=0.408982).

The pulse edge steepness in Fig. 3(b) relies on both chromatic dispersion and TOD, as a result the individual contribution of TOD is unclear. Besides, since the DSR in the case of $D > 0$ and $\beta_3 = 0$ is already highly steep, the variation of pulse edge rise time caused by TOD cannot be clearly resolved in Fig. 3(b). Here, we use the relative change of rise time $\eta_{rise\_time}$ to quantitatively describe the impact of TOD on pulse edge steepness. It is defined by

$$\eta_{rise\_time} = \frac{T_{risetime\_withTOD} - T_{risetime\_withoutTOD}}{T_{risetime\_withoutTOD}} \quad (2)$$

where $T_{risetime\_withTOD}$ and $T_{risetime\_withoutTOD}$ are the rise times of pulse edge with $\beta_3 = 0$ and $\beta_3 \neq 0$, respectively. $\eta_{rise\_time} > 0$ ($< 0$) corresponds to steepen (stretch) pulse edge, and its absolute value reflects the relative change of rise time. Figure 3(c) shows the calculated $\eta_{rise\_time}$ of Fig. 3(b) with $\beta_3 = 0.015$, and Fig. 3(d) shows $\eta_{rise\_time}$ of leading pulse edge with three different TOD of $\beta_3 = 0.005, 0.010, 0.015$. In both plots, $\eta_{rise\_time}$ varies as a function of chromatic dispersion and shows two peaks: one is at $D \approx -0.96$ and the other is at $D \approx 0.60$. When chromatic dispersion deviates from the two peaks, the absolute value of $\eta_{rise\_time}$ decreases gradually, indicating the weakened contribution of TOD on pulse edge steepness. Besides, $\eta_{rise\_time}$ in normal chromatic dispersion region is relatively higher when compared to the anomalous region. $|\eta_{rise\_time}|$ reaches the maximum value at $D \approx -0.96$. In the cases of $D \approx 0$ or $D \gg 0$, $\eta_{rise\_time}$ approaches to zero, indicating the less sensitivity of pulse edge steepness on TOD.

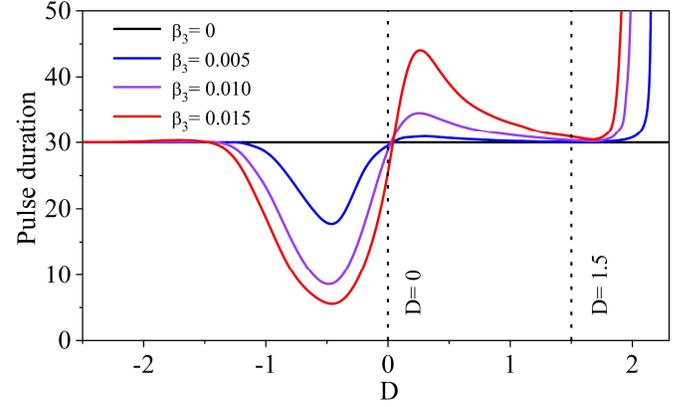

Fig. 5. Pulse duration as a function of chromatic dispersion at different $\beta_3$. The curve with $\beta_3$=0 has a constant pulse duration of 30, and other curves are calculated based on this curve but taking TOD into consideration.

Figure 3 shows that, if one of the pulse edges is stretched, the other will be steepened. And in Fig. 2(b), the steepness of leading edge is continuously enhanced when $\beta_3$ increases from 0 to 0.24. These results seemingly imply that the DSR with an infinite steep pulse edge could be formed if TOD is strong enough. However, our numerous simulation results show that the pulse edge cannot be infinitely steepened by TOD, and both pulse edges turns to be stretched when $\beta_3$ exceeds a certain value. In order to clearly illustrate this, here we simulated the rise times of leading and trailing edges as a function of $\beta_3$ at two different parameter values of ($D$=1.5, $\varepsilon$=0.44) and ($D$=0, $\varepsilon$=0.408982), as seen in Figure 4(a) and 4(b), respectively. In Fig. 4(a), the rise times are $T_{risetime\_leading\_edge} = T_{risetime\_trailing\_edge} = 2.57$ at $\beta_3 = 0$. As $\beta_3$ increases from 0 to 0.2, $T_{risetime\_leading\_edge}$ always decreases, but $T_{risetime\_trailing\_edge}$ first decreases to a minimum of ~2.43 at $\beta_3 = \sim0.14$ and then starts to increase, suggesting that the pulse edge steepness cannot be infinitely enhanced by TOD. Figure 4(b) shows a similar phenomenon to Fig. 4(a). However, in this case, because the pulse edges at $\beta_3 = 0$ have a shorter rise time of ~2.17 and thus higher steepness, the leading pulse edge is only weakly steepened and then being continuously stretched, as seen the inset of Fig. 4(b).

C. *The pulse duration of DSR in the space of ($\beta_3$, D, $\varepsilon$)*

The pulse duration also depends on both TOD and chromatic dispersion. In order to illustrate this, the pulse duration as a function of $D$ is plotted in Fig. 5 with different TOD ($\beta_3 = 0$, 0.005, 0.010, and 0.015). In this plot, the horizontal straight line without TOD corresponds to constant pulse duration of $T_{pulse\_duration}$ =30.0, and all other curves are simulated by adding different TOD to this curve. These curves show that, with constant TOD, the pulse duration strongly relies on chromatic dispersion. Depending on the value of chromatic dispersion, the pulse duration becomes either shorter or longer when TOD is considered. In large normal dispersion region ($D < \sim -1.5$), the pulse duration ($T_{pulse\_duration}$) is slightly increased by TOD. Decreasing of $T_{pulse\_duration}$ occurs in the range of $\sim -1.5 < D < 0.05$. The most significant drop in pulse duration occurs at $D = -0.46$, where the pulse duration falls down from 30 to 5.6 as $\beta_3$ increases from 0 to 0.015. When $D > 0.05$, TOD always leads to longer pulse. In the vicinity of $D \approx 0.26$ as well as the region of $D \gg 0$, TOD leads to a dramatic increase of the pulse duration. This trend is particularly remarkable in the large anomalous dispersion region ($D \gg 0$).

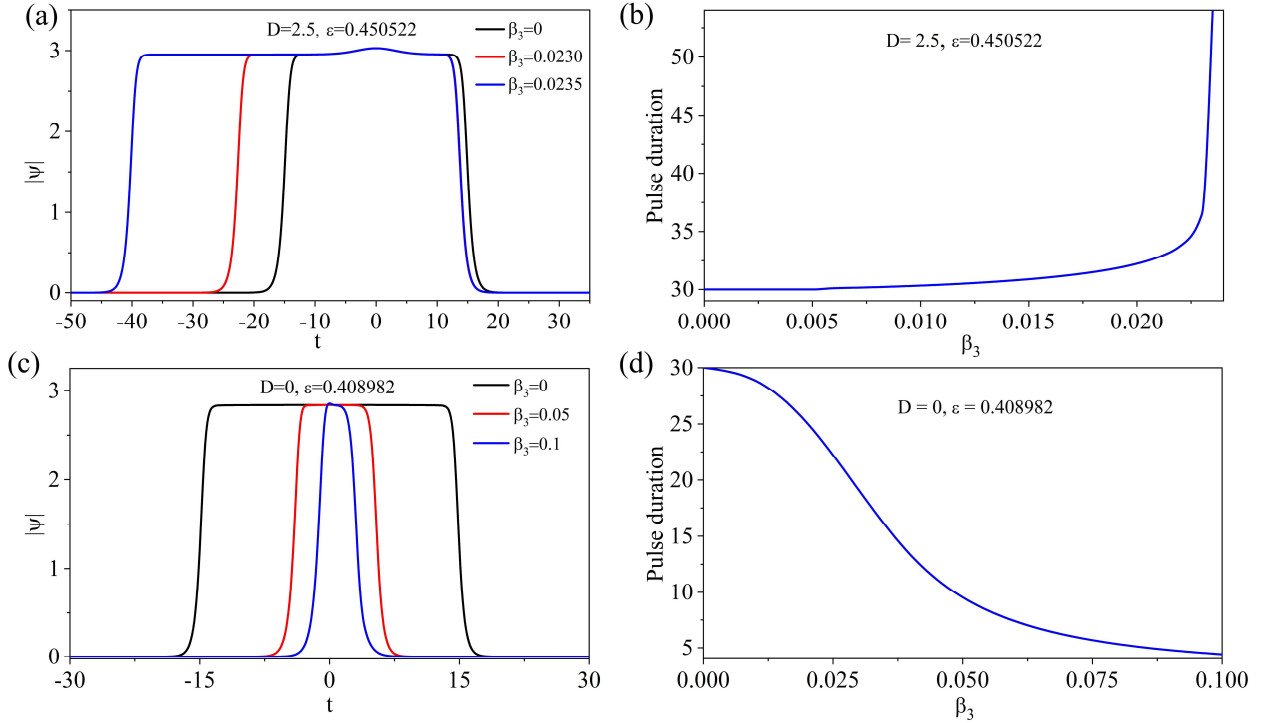

FIG. 6. (a) and (c) are pulse profiles for different TOD at two typical cases of ($D$=2.5, $\varepsilon$=0.450522) and ($D$=0, $\varepsilon$=0.408982), respectively. (b) and (d) are corresponding evolution of pulse duration $T_{pulse\_duration}$ as a function of $\beta_3$.

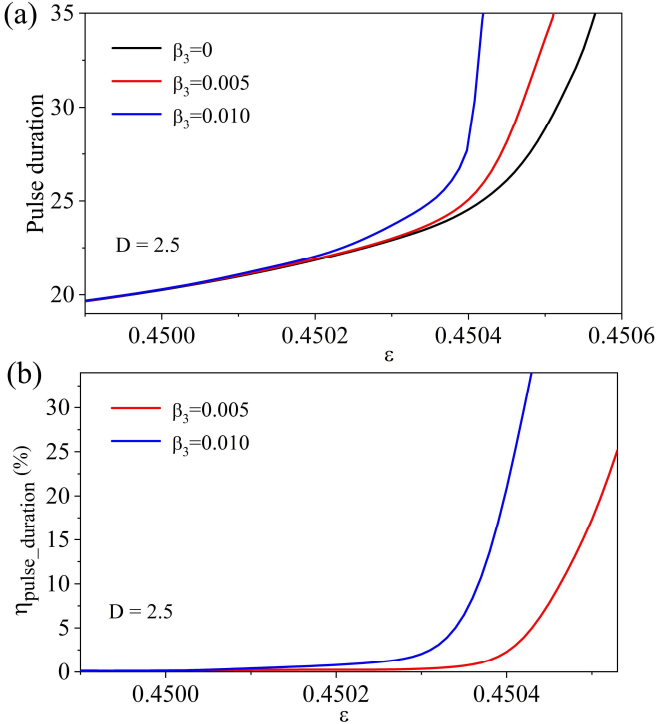

FIG. 7. (a) Pulse duration as function of nonlinear gain $\varepsilon$ under three different TOD values of $\beta_3 = 0$, 0.005, and 0.010. (b) The relative change of pulse duration ($\eta_{pulse\_duration}$) in term of TOD.

Figure 5 also indicates that, with constant chromatic dispersion D, the pulse duration shows either a monotonous increase or decrease. For better clarity, two typical results are presented in Fig. 6. Figure 6(a) shows the pulse profiles of DSR with different TOD ($\beta_3 = 0$, 0.023, and 0.0235) at the point of ($D$=2.5, $\varepsilon$=0.450522), and Figure 6(b) shows the corresponding evolution of pulse duration as a function of $\beta_3$. In this case, $T_{pulse\_duration}$ rapidly increases from 30.0 to 54.1 when $\beta_3$ slightly increases from 0 to 0.0235. A typical feature of Fig. 6(b) is that the rate of increase in pulse duration speeds up as $\beta_3$ increases. Remarkably, when $\beta_3 > \sim 0.023$, $T_{pulse\_duration}$ rises sharply as a function of $\beta_3$. Similar trend is also presented from Fig. 2(b). When TOD leads to the decrease of pulse duration, the situation becomes slightly different. Figure 6(c) and 6(d) shows the simulation results at ($D$=0, $\varepsilon$=0.408982), where $T_{pulse\_duration}$ continuously decreases from 30.0 to 4.4 when $\beta_3$ increases from 0 to 0.1.

In the above analysis, the role of TOD is only investigated for DSR with pulse duration $T_{pulse\_duration} = 30.0$. However, the pulse duration of DSR is proportional to nonlinear gain $\varepsilon$. When $\varepsilon$ is small enough, the DSR will evolve into dissipative soliton regime, which is characterized by a Gauss pulse profile and small pulse energy relative to DSR [7]. Therefore, it is important to know how the TOD alters the relationship between pulse duration and nonlinear gain. Figure 7(a) shows the evolution of the pulse duration $T_{pulse\_duration}$ as a function of nonlinear gain $\varepsilon$ under three different TOD ($\beta_3 = 0, 0.005,$ and $0.010$), where $D$ is set to 2.5. In this plot, regardless of the value of TOD, $T_{pulse\_duration}$ continuously increases with the increase of $\varepsilon$, and the increase speed ($\partial T_{pulse\_duration}/\partial \varepsilon$) is accelerated gradually. The pulse duration at $\beta_3 \neq 0$ is always longer than the case of $\beta_3 = 0$, and higher $\beta_3$ leads to longer pulse duration. The change ratio of pulse duration in term of TOD is described by using:

$$\eta_{pulse\_duration} = \frac{T_{pulseduration\_withTOD} - T_{pulseduration\_withoutTOD}}{T_{pulseduration\_withoutTOD}} \quad (3)$$

where $T_{pulseduration\_withTOD}$ and $T_{pulseduration\_withoutTOD}$ are pulse duration of soliton solution with $\beta_3 = 0$ and $\beta_3 \neq 0$, respectively. Figure 7(b) shows the variation of $\eta_{pulse\_duration}$ as a function of $\varepsilon$ under different TOD, where $\eta_{pulse\_duration}$ increases with the increase of $\varepsilon$, indicating that the pulse duration is increasingly dependent on TOD as nonlinear gain increases.

## IV. Conclusion

In conclusion, we numerically investigate the impact of TOD on DSR in both anomalous and normal chromatic dispersion regions. The simulations demonstrate that TOD is an important factor for the pulse duration and thus pulse energy of DSR, regardless of the sign of TOD. In some chromatic dispersion region, TOD leads to rapid decreasing of pulse duration. However, when chromatic dispersion is appropriately set, TOD leads to a longer pulse duration and higher pulse energy. Particularly, in large anomalous dispersion region, even weak TOD can leads to dramatic increase in pulse duration. This perhaps answers the frequent reports of DSR generation based on mode-locked fiber lasers consisting of long anomalous chromatic dispersion fibers (hundred meters), which leads to the accumulation of a large amount of TOD [2], [6]. Under constant chromatic dispersion, the pulse duration is increasingly dependent on TOD as the nonlinear gain increases. Another notable effect is that TOD leads to an asymmetric pulse profile of DSR. As a result, the leading and trailing pulse edges show different steepness. The edge steepness caused by TOD varies as a function of chromatic dispersion. In the vicinity of near zero anomalous chromatic dispersion, the DSR shows highly steepened pulse edges which is insensitive on TOD effect, but the pulse duration can be remarkably longer if TOD is introduced, which facilitates the generation of DSR with higher pulse energy and steeper pulse edges.

**Yazhou Wang** received the B.Sc. degree in physics from Southwest University, the M.Sc. degree in 3D displaying from Sichuan University, and the Ph.D. degree in nonlinear fiber optics from the University of Electronic Science and Technology of China. During his Ph.D., he was a Visiting Student with the Technical University of Denmark in 2018, with the study on air plasma based THz generation. He received the Ph.D. degree in 2019, he joined Technical University of Denmark as a Post-Doc. His current study focuses on infrared gas-filled fiber Raman laser design and applications in gas detection. He is a member of OSA and SPIE society.

**Abubakar Isa Adamu** (Member, IEEE) received the B.Sc. degree in Engineering physics from Gaziantep University, Turkey and the M.Sc. degree in Material Science and Nanotechnology from Bilkent University, Turkey, where he worked on development of *Artificial Nose*, for sensing of toxic gases with National Nanotechnology Research Centre (UNAM), Turkey. He was a Visiting Student with the University of Chicago, USA, in 2013, where he learnt therapeutic modalities in radiation oncology. He was a Guest Research Student with the University of Twente, The Netherlands, in 2015, where he worked on membranes for next-generation EUV lithography. . He received the Doctorate degree in 2020 from the Technical University of Denmark in Photonics Engineering, where we worked on hollow-core optical fibres, ultrafast nonlinear optics, supercontinuum laser and fabrication of hybrid speciality fibres. He is currently a Manufacturing Engineer with Lumenisity Ltd, U.K. Dr. Adamu is a member of OSA, EPS, and SPIE. He was recently awarded the Outstanding Student Paper Award at SPIE Photonics West, 2019, San Francisco, USA and Best Research Poster Award by Danish Optical Society (DOPS) and nominated. Ambassador of Optica 2021 (formerly Optical Society of America).

**Lujun Hong** received the B.Sc. degree in optical information science and technology, and the M.Sc. Degree and Ph.D. degree from Nanchang University, Nanchang, China, respectively. During the Ph.D., he was a Visiting Student with the Technical University of Denmark in 2018, with the study on THz generation and detection based on high-energy femtosecond laser. He received the Ph.D. degree in 2020 and joined Nanchang University as a lecturer from 2020 to 2022. He is currently a Postdoc with Technical University of Denmark. His currently research areas include micro-plastic detection based on gas-filled hollow core fiber Raman laser, broadband THz spectroscopy based on ultra-fast lasers, and unidirectional propagation.

**Md. Selim Habib** (Senior Member, IEEE) received the B.Sc. and M.Sc. Degrees in electrical and electronic engineering from the Rajshahi University of Engineering and Technology, Rajshahi, Bangladesh, in 2008 and 2012, respectively. He received the Ph.D. degree from Technical University of Denmark (DTU) in 2017. After finishing his Ph.D., he joined as a Postdoctoral Researcher in Fibers Sensors and Supercontinuum Group with the Department of Photonics Engineering, DTU. After finishing his Postdoctoral Fellowship with DTU, he worked as a Postdoctoral Research Associate with CREOL, The College of Optics and Photonics, University of Central Florida, USA, from 2017 to 2019. He is currently an Assistant Professor of Electrical and Computer Engineering with Florida Polytechnic University, USA. He has authored or coauthored more than 50 articles in referred journals. His research mainly focuses on design, fabrication, and characterization of low loss hollow-core fiber in the near-IR to mid-IR, light gas nonlinear interaction in hollow-core fibers, supercontinuum generation, and multi-mode nonlinear optics.

Dr. Habib is a Senior Member of Institute of Electrical and Electronics Engineers (IEEE), Optical Society of America (OSA) Early Careers Member, and Executive officer of OSA Fiber modeling and Fabrication group. Dr. Habib is an Associate Editor of IEEE ACCESS, and Feature Editor of Applied Optics (OSA). He was the recipient of the University Gold Medal Award from Rajshahi University of Engineering and Technology in 2014.

**Mengqiang Cai** received the B.Sc. degree in optical Information sciences and technology from Hefei University of Technology, the M.Sc. degree in optics from Nankai University, and the Ph.D. degree in optics from Nankai University. He received the Ph.D. degree in 2017, he joined Nanchang University. His currently study focuses on terahertz optical field manipulation and detection.

**Jie Xu** received the B.S. degree in applied physics and M.S degrees in plasma physics from Nanchang University and the Ph.D. degree in material science and engineering from Nanchang University in 2020. His research interests include nonreciprocal electromagnetic propagation based on magneto-optical materials and subwavelength functional devices. From 2020 to now, he is a teacher of Southwest University, Luzhou, China.

**Jianfeng Li** received the B.S. degree in applied physics from the Sichuan University, Chengdu, China, in 2003, and the M.S. and Ph.D. degrees in optical engineering from the Institute of Nano-Optics, Department of Physics, Sichuan University, China, in 2005 and 2008, respectively. In 2008, he joined the School of Optoelectronic Information, University of Electronic Science and Technology of China, Chengdu, China, where he became an Associate Professor in 2009. In 2011, he joined the Centre for Ultrahigh Bandwidth Devices for Optical Systems (CUDOS), University of Sydney, Australia, as a Visiting Scientist. In 2013, he joined the Aston Institute of Photonic Technologies, Aston University, U.K. as a Marie Curie IIF Fellow funded by the European Commission's Seventh Framework Program (FP7). His current research interests include fiber lasers, fiber sensors, and nonlinear fiber optics.

**Christos Markos** Christos Markos received his B.Eng. (Hons) in 2007 and M.Sc. (with Distinction) degree in 2008 from University of Liverpool in Electrical Engineering and Electronics. He continued his studies as PhD student at the National Hellenic Research Foundation, Theoretical and Physical Chemistry Institute, Athens and received his Ph.D. in Optics/Optoelectronics in 2013. His main research activities are broad within the field of photonics and particularly in gas-filled hollow-core fiber lasers, chalcogenide glasses and multimaterial fiber optics towards development of novel neural interfaces and optoelectronic smart devices. Christos has joined and worked with several distinguished research groups in USA and Europe including the Multi-material Optical Fiber Devices Group in College of Optics and Photonics (CREOL), USA and Mid-Infrared Photonics Group in the University of Nottingham, UK among others. He has established three state-of-the-art laboratories at DTU Fotonik for soft glass synthesis, extrusion and fiber fabrication and two more labs dedicated for his Neurophotonics activities. Christos holds the position of Associate Professor, and he is the head of the Neural Devices and Gas Photonics group at DTU Fotonik. He is a member of OSA, SPIE, IEEE societies and co-founder of NORBLIS ApS.